\documentclass{ws-procs975x65}
\begin{document}
\title{Stability in Generalized  Modified  Gravity  }

\author{Guido Cognola$^{a}$, Lorenzo Sebastiani$^{a}$, Sergio Zerbini$^{a}$}

\address{$^a$Dipartimento di Fisica, Universit\`{a} degli Studi di Trento,\\
Trento, 38100, Italy\\
INFN - Gruppo Collegato di Trento\\
}



\begin{abstract}
The stability issue of Generalized  modified gravitational models is 
discussed with  particular emphasis to de Sitter solutions. Two approaches 
are briefly presented.
   
\end{abstract}


\bodymatter



\section{The de Sitter stability issue} 
It is well known that recent astrophysical data are in agreement with a 
universe in current phase of accelerated expansion, in contrast with the predictions of Einstein gravity in FRW  space-time.
Most part of energy contents, roughly $75\, \%$ in the universe is due to  mysterious entity  with negative 
pressure: Dark Energy. The simplest explanation is  Einstein gravity plus a small positive cosmological constant.
As an  alternative, one may consider more drastic modification of General Relativity: Modified Gravity Models, 
see for example \cite{Capo,turner,No,fara}.

We shall deal with modified generalized models, described by a Lagragian density $F(R,P,Q)$ \cite{easson}, 
where $R$ is the Ricci scalar, and $P= R_{\mu \nu}R^{\mu \nu}$, and  
$Q=R_{\mu \nu \alpha \beta}R^{\mu \nu \alpha \beta}$ are quadratic curvature invariants. In particular  
the Gauss-Bonnet topological invariant reads $G=R^2-4P+Q$. 
The stability of the de Sitter solution, relevant for Dark energy, 
 may be investigated in these Gauss-Bonnet models in several ways. 
We limit ourselves to the following  two approaches: Perturbation of Esq. 
of Motion  in the Jordan frame and  Dynamical System Approach in 
FRW space-time.  

\section{Stability  of $F(R,P,Q)$ model in the Jordan frame} 
The starting point is the trace of the equations of motion, which is 
trivial in Einstein gravity $R=-\kappa^2 T$, but, for a  general $F(R,P,Q)$ model, reads
\begin{equation}
\nabla^2 \left(3F'_R+RF'_P \right)+2\nabla_\mu \nabla_\nu \left[ \left(F'_P+2F'_Q\right)R^{\mu \nu}\right]-2F+RF'_R
+2\left(F'_P+F'_Q\right)=\kappa^2 T\,.
\end{equation}
Requiring $R=R_0=Cts $, $P_0=Cts $, and $Q_0=Cts $ one has de Sitter existence condition in vacuum  
\begin{equation}
2F_0-R_0F'_{R_0}-2P_0F'_{P_0}-2Q'_{Q_0}=0 \, .
\end{equation}
Perturbing around  dS space, namely   $R=R_0+\delta R $,  and  with $P=P_0+\delta P $, and $Q=Q_0+\delta Q $, 
observing that 
$\delta P= \frac{R_0}{2} \delta R $, and $\delta Q= \frac{R_0}{6} \delta R $, one arrives at the perturbation Eq.
\begin{equation}
-\nabla^2 \delta R + M^2 \delta R =  0 \, ,
\end{equation}
in which the scalaron effective mass reads
\begin{equation}
M^2= \frac{R_0}{3}\left(\frac{F'_{R_0}+4H_0^2
\left(\frac{3}{2}F'_{P_0}+F'_{Q_0} \right)}{R_0 [A_R+A_Q+A_P+4H_0^2 \left(\frac{3}{2}F'_{P_0}+F'_{Q_0}\right)]}-1\right)
\end{equation}
where
\begin{equation}
A_R=F''_{R_0 R_0}+6 H^2_0F''_{R_0 P_0}+4H_0^2F''_{R_0 Q_0}
\end{equation}
\begin{equation}
A_Q=2H^2_0\left(F''_{R_0 P_0}+6 H^2_0F''_{P_0 P_0}+4H_0^2F''_{P_0 Q_0}\right)
\end{equation}
\begin{equation}
A_P=4H_0^2\left( F''_{R_0 Q_0}+6 H^2_0F''_{Q_0 P_0}+4H_0^2F''_{Q_0 Q_0}\right)
\end{equation}

Thus, if  $M^2>0$, one has  stability of the dS solution. In the particular case $F(R,G)$, one has \cite{Cogno07,Cogno08} 
\begin{equation}
\frac{9F'_{R_0}}{R_0[9F''_{R_0 R_0}+6 R_0F''_{R_0 G_0}+R_0^2F''_{G_0 G_0}] }> 1\,.
\end{equation}
In the case of a $F(R)$ models, one has the well known condition $\frac{F'_{R_0}}{R_0F_{R_0}'' }> 1 $.
\section{ Dynamical System Approach}
This approach has been used by many authors. One works in a cosmological setting, namely with a  FRW metric, and the main 
idea consists in rewriting  the generalized Einstein-Friedman equations in an 
equivalent system of first order differential equations, introducing
 new dynamical variables $\Omega_i$
\begin{equation}
\frac{d}{d t} {\vec \Omega(t)}=\vec v(\vec \Omega(t))\,.
\end{equation}
Here the evolution parameter has been denoted by $t$. 
The critical (or fixed) points are defined by $ \vec v(\vec \Omega_0)=0$.
The key point is: 

{\bf Hartman-Grobman theorem}: 
{\it The orbit structure of a dynamical system  in the neighbourhood of a 
hyperbolic fixed point is topologically equivalent to the orbit 
structure of  the associated linearized dynamical system,
 defined by a stability matrix $M_0$}.
\\ Thus, in order to study the stability of the above 
non linear system of differential Eqs. at critical points, it is sufficient 
to investigate the  related linear system of differential Eqs.:
\begin{equation}
\frac{d}{d t} \delta {\vec \Omega(t)} =M_0 \delta \vec \Omega(t)\,,
\quad   M_0 \quad \mbox{Jacobian matrix} 
 \quad \mbox{evaluated at } \vec \Omega_0
\end{equation}
The solution of the linearization is well known and the evolution is determined by the signs of the eigenvalues of $ M_0$. As a result, the non linear system is stable if all eigenvalues of 
the matrix $M_0$ have negative real parts.

As an example, let us consider a modified model  $R+f(G)$. The related autonomous system in the two 
unknown quantities $G$ and $H$ reads
\begin{equation}
\dot G=\frac{1}{24 f''_GH^3}\left((Gf'_G-f)-6H^2\right)\,, \quad
\dot H=\frac{G}{24 H^2}- H^2 \,. 
\end{equation}
The critical points are defined by $\dot G=0$ and $\dot H=0$. Thus,  we have the solutions $24 H_0^4=G_0$ and 
$G_0f'_0-f_0=6H_0^2$ and these correspond to a de Sitter critical point  with Gauss-Bonnet invariant. 
The linearized system around de Sitter critical point reads
\begin{equation}
\dot \delta G=H_0 \delta G-\frac{1}{2H_0^2f_0''} \delta H \,,\quad
\dot \delta H=\frac{\delta G}{24 H_0^2}-4 H_0 \delta H \,. 
\end{equation}
One can read off the stability matrix and the stability condition is
$\frac{9}{R_0^3 f_0''} >1$, in agreement with the previous approch.

\section{Concluding remarks}  
Modified gravity may be seen as the phenomenological description 
of a fundamental unknown theory.
From this point of view, corrections to 
Einstein-Hilbert action depending on higher order  curvature invariants are 
likely to be expected (Lovelock gravity is an example). 

Among many existing approaches, two methods have been illustrated in 
order to investigate the stability of
these models around de Sitter critical points, and  the dS stability 
conditions has been derived within  two possible approaches.  

These methods have owns advantages and problems, and, in our opinion, 
both  permit to study
critical points and stability for modified gravitational 
models depending on arbitrary geometric invariants, 
generalising the results obtained for $F(R)$ models with other methods 
(see, for example \cite{Cogno05,Cogno06,zu}).
We conclude noting that the dynamical system approach has also been applied to non local $F(R)$ 
models \cite{bb,bbb}.  




\end{document}